Intermittency in Voyager's Magnetic Field Beyond the Heliosphere

L. Y. Khoo ,[1] G. Livadiotis ,[1] D. J. McComas ,[1] M. E. Cuesta ,[1] and J. S. Rankin [1]

[1]*Department of Astrophysical Sciences, Princeton University, Princeton, NJ 08544, USA*

## ABSTRACT

As the two Voyager spacecraft traveled beyond the heliosphere, they encountered a magnetic field environment that had never been observed before. Studies have attempted to characterize this new regime by examining the magnetic field intermittency. This is typically done by fitting the optimal kappa distribution function and interpreting its so-called q-statistics to characterize the magnetic field increments. Using this approach, recent findings concluded that beyond a certain distance, the magnetic field increments in the very local interstellar medium (VLISM) follow Gaussian statistics, unlike those both inside the heliosphere and in the region just beyond the widely accepted heliopause location, raising questions about the heliopause identification. This study explores this issue in detail by (1) optimizing the derivation of the distribution function, (2) examining whether and how the results depend on increment windows and time periods, and (3) determining the statistical behavior of the examined time series. Using magnetic field measurements from Voyager 1, we present two independent techniques and introduce a statistical framework to systematically analyze the distributions of magnetic field increments. Contrary to previous findings, we find that magnetic field increments in the VLISM do not follow a Gaussian distribution ($\kappa \rightarrow \infty$) and instead are in the non-Gaussian range of kappa values (3-7, when analyzed on a 30-day statistical period). We further demonstrate how erroneous, statistically induced results can arise that mimic Gaussian-like results when mixing different structures in such analyses. Our results show that Voyager 1 still travels in the intermittent magnetic field environment of the VLISM.

## 1. INTRODUCTION

As the Voyagers explore the Very Local Interstellar Medium, they continue to uncover unexpected behaviors in particles, plasma and magnetic fields in the region beyond the heliopause, which is modified by the presence of the heliosphere. Analysis of these unprecedented measurements has led to a variety of interpretations of this unique data set. Here, we develop a statistical framework to gain new insights into the behavior of the VLISM magnetic field by analyzing the magnetic field increments.

Intermittency often indicates strong spatially localized fluctuations of a physical random variable. If the fluctuations exhibit a Gaussian distribution, this suggests decoherence and structureless fluctuations in the system. In contrast, a non-Gaussian distribution is often described within the framework of q-statistics, indicating the existence of coherence in the system. In q-statistics, stable distributions are called q-Gaussian (Tsallis 2009), and they have an equivalent form and statistics to kappa distributions (Livadiotis & McComas 2009). Here the term 'coherence' indicates a connection or correlation between various fluctuations examined in the same distribution, such as when there are different but connected fluctuation sources; conversely, the opposite term, 'decoherence', indicates independent stochastic fluctuations. This was also the rationale behind the intermittency characterization where a random variable, described by q-statistics, was thought to be an evolving process that would eventually be described by Gaussian statistics (non-intermittent). Nevertheless, q-statistics is a rather stable statistics, complete with a generalized central limit theorem where the distributions of fluctuations converge to kappa distributions (Umarov et al. 2008). Despite the connection between the q-statistics and intermittency, it is not necessary that the distributions will evolve towards, or further away from, Gaussian behavior.

Spatially intermittent structures, described by q-statistics, can arise from various forms such as current sheets, discontinuities, and shock waves. Numerous authors have studied magnetic field intermittency in the heliosphere in order to investigate, among other things, turbulence under various solar wind conditions (e.g., Bruno et al. 2001; Greco et al. 2008; Chhiber et al. 2021; Cuesta et al. 2022), the transition of q-statistics to normal, Gaussian statistics (e.g., Livadiotis & McComas 2011a), detection and characterization of heliospheric structures (e.g., Zirnstein & McComas 2015; Richardson et al. 2022; Burlaga et al. 2023), and the connection of intermittency to the thermodynamic behavior of solar energetic particles (Cuesta et al. 2025, submitted).

As the most distant spacecraft in human history, the two Voyager spacecraft have enabled studies that examine magnetic field intermittency in the heliosheath region (Burlaga et al. 2006a) and, more recently, in the very local interstellar medium (VLISM; Burlaga et al. (2022, 2024a,b)). Burlaga et al. (2024b) examined the intermittency of magnetic fields from Day 1-271 of 2023 by fitting the sequential magnetic field increments using q-statistics in linear space and concluded that the magnetic fields measured by Voyager during this period follow a Gaussian distribution. This suggests that Voyager may have entered a new region, and the timing of this transition generally coincides with the predicted location of the heliopause by Fisk & Gloeckler (2022), raising questions about the widely accepted previously defined heliopause location based on solar wind plasma (Gurnett et al. 2013).

Inspired by their work, we examine the magnetic field intermittency, specifically the sequential magnetic field increments, during the first 271 days of 2023 (consistent with what was investigated in Burlaga et al. 2024b) using two different kappa techniques. The aims of this study are twofold: (1) to examine if the result varies when the fitting is performed in logarithmic space, which is more sensitive to a distribution's tail; (2) to investigate if the kappa value depends on the increment window and statistical period. This work is organized as follows: Section 2 explains the q-statistic theory and its implications for intermittency and incoherence. Section 3 provides information about the Voyager data used in this study. Section 4 describes the two techniques we used to determine the kappa values, and Section 5 presents the results of the kappa techniques. Section 6 discusses and summarizes the key findings of this work and its implications.

## 2. BACKGROUND: Q-STATISTICS AND KAPPA DISTRIBUTION

Non-extensive statistical mechanics has provided a very successful generalization of the Boltzmann-Gibbs statistical mechanics, which describes a system in the classical adaptation of thermal equilibrium where there are no correlations among the particles. By maximizing the generalized non-extensive entropy consistent with thermodynamics (Tsallis 1988; Livadiotis & McComas 2023) (also known as Tsallis entropy), one can derive the Tsallis q-exponential distribution, a canonical distribution that is parameterized by q-index and describes the average behavior of a particle system that may or may not be at thermal equilibrium (e.g. Tsallis 2009; Livadiotis 2014).

Meanwhile, the kappa distribution has been widely used in the space physics community to describe the non-Gaussian nature of space plasmas since 1966 (see Livadiotis 2017, and references therein). These distributions are labeled and parameterized by the flexible parameter kappa, $\kappa$, that we now know characterizes the thermodynamic stationary state of the plasma. The relationship between Tsallis' q-index and the thermodynamic kappa has been established, namely $q = 1 + 1/\kappa$ (Livadiotis & McComas 2009). The connection of the kappa distribution to the solid foundation of non-extensive statistical mechanics lays the groundwork for many further developments in the field. One of the important developments involves linking kappa to the characterization of thermodynamic processes (Livadiotis & McComas 2012; Livadiotis 2019; Fichtner et al. 2021). By enabling the use of kappa distribution to describe temperature for systems residing in stationary states out of the classical thermal equilibrium, it also establishes kappa as an independent thermodynamic parameter like temperature and density, to characterize non-equilibrium systems (Livadiotis & McComas 2012).

Theoretically, the 3-dimensional kappa value ranges from 1.5 to $\infty$. The higher the kappa value, tending to infinity, the closer the distribution to a Gaussian distribution, and the less correlated the particles are inside the system. For a distribution that is closer to the other extreme (1.5), it suggests that the system is farthest away from equilibrium, or 'anti-equilibrium' (Livadiotis & McComas 2013), and the more correlated the particles are within the system.

However, the kappa distribution or q-exponential is not limited to describing the energies of plasma populations; it can be applied to any physical random variable that is subject to correlation. When the kappa distribution is used to characterize the statistical distribution for the time evolution of a certain parameter such as the magnetic field increments (e.g., Greco et al. 2008; Cuesta et al. submitted), the physical meaning of the kappa index is preserved. This is crucial because it means that the kappa value can provide insights on the characteristics of these increment distributions, such as the correlation within the system, and thus offering clues on the intermittency and even identifying different non-equilibrium systems or structures in space.

## 3. DATA: VOYAGER MAGNETIC FIELD MEASUREMENTS

This study uses 48-second average magnetic field measurements from the Voyager-1 Magnetic field experiment (MAG; Behannon 1977) that are available in the radial, tangential and normal (R, T, and N) magnetic field vectors as well as its magnitude from three field components. R refers to the sun-to-spacecraft vector, T is the cross product of the solar rotation axis (northward) and R, and N completes the right-handed coordinate system (Franz & Harper 2002). Recent studies have verified that vectors T and N are generally well-calibrated and agree with models of the magnetic field in the outer heliosphere (Izmodenov & Alexashov 2020; Rankin et al. 2023) while the R component appears to have a larger discrepancy with the models. The T and N vectors were calibrated using inputs from spacecraft spinning maneuvers such as magnetometer roll maneuvers (also known as MAGROLs), which cannot be used to calibrate the radial vector. Therefore, $B_R$ was "calibrated" using the average expected value based on the Parker spiral inside the heliosphere (Berdichevsky 2009), and with less robust methods outside the heliosphere (Berdichevsky 2016). Nevertheless, in this study we use all three vector components ($B_N$, $B_T$, $B_R$) as well as the magnitude ($B$) that are from the publicly available data hosted at CDAWeb: https://cdaweb.gsfc.nasa.gov. We also use $B_{TN}$ that is determined from $\sqrt{B_T^2 + B_N^2}$.

## 4. METHODOLOGY: DETERMINING KAPPA PARAMETERS OF MAGNETIC FIELD INCREMENTS

Magnetic field increments and their statistical distribution are useful tools to examine the intermittency in various space environments, including the heliosheath and VLISM (Burlaga et al. 2006a,b, 2024a). Past studies have demonstrated the use of q-index to characterize the statistical distribution of the magnetic field increments beyond the heliosphere (Burlaga et al. 2020). Given the equivalent relationship between the q-index and kappa index, and the extensive history of the broad usage of kappa distributions in space physics studies, we are motivated to use the kappa index for such characterizations in this study.

This section provides description of key parts of our analysis: (1) how to compute magnetic field increments, and (2) how to determine the parameter kappa from the statistical distribution of the magnetic field increments for a specific period.

### 4.1. *Computing the magnetic field increments*

Figure 1 summarizes the steps involved to compute the sequential magnetic field increments in this study. Like Burlaga et al. (2024b), this study utilizes Voyager 1 magnetic field data from the beginning of Day 1 to the end of Day 271 of year 2023. All analysis in this study begins with the 48-second averaged magnetic field measurements. We further average the 48-second averaged magnetic field data over longer periods ($\bar{B}_i(t)$; ranging from 1 minute to 200 minutes) in non-overlapping samples from which we compute sequential magnetic field increments, such that the increment scale is exactly equal to the resampling duration. If there is no data within the resampling duration, $\bar{B}_i(t)$ is assigned as NaN.

Following the notation used by Burlaga et al. (2024b), sequential magnetic field increments are defined as:

$$\delta B_i(t, \bar{\tau}) = \bar{B}_i(t + \bar{\tau}) - \bar{B}_i(t) \tag{1}$$

where

$$\bar{B}_i = \langle B_i(t) \rangle_{\bar{\tau}} \tag{2}$$

is the average of the magnetic field component or magnitude $B$, t is the observation time, i is each component of magnetic field (in this study, we use R, T, N, and NT), and $\bar{\tau}$ is both the increment scale and resampling duration.

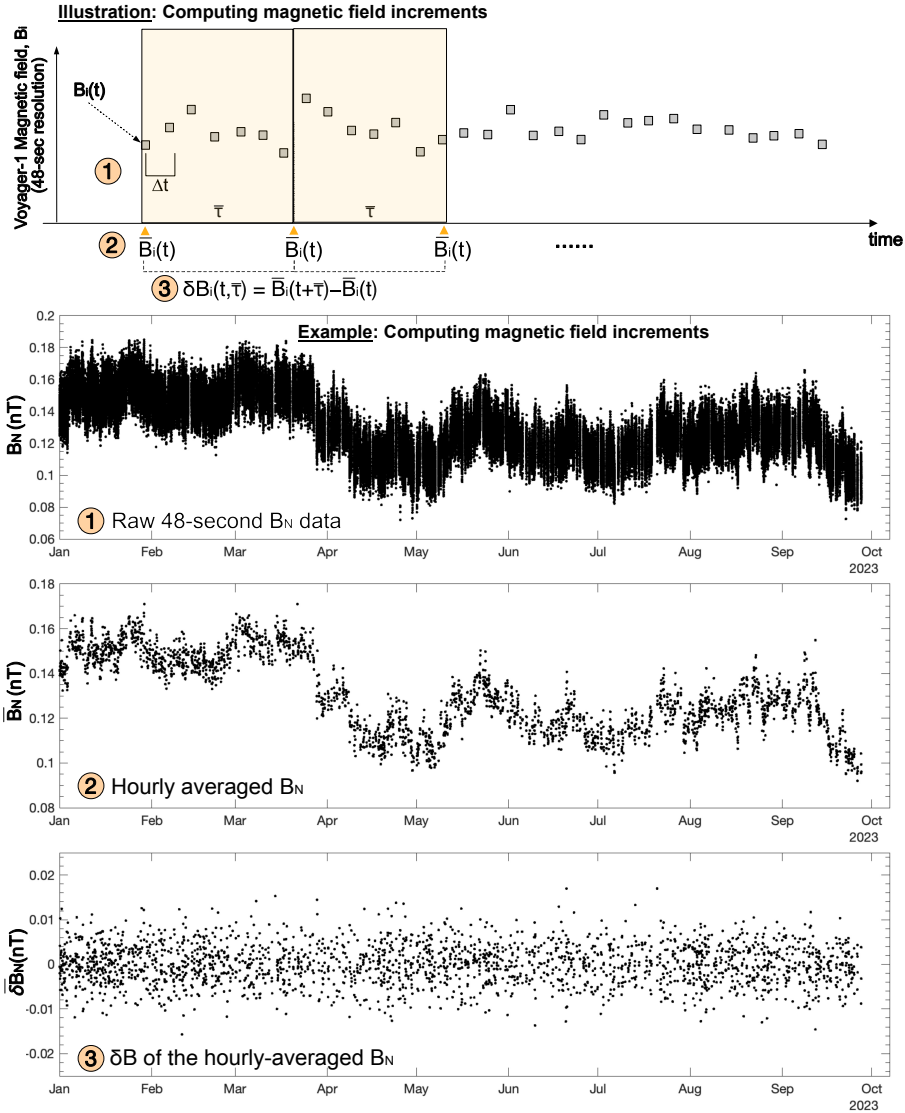

*Figure 1. (Top) Illustration of the steps involved in computing the sequential magnetic field increments. (Panels 2-4) Application of steps to compute the sequential magnetic field increments of $B_N$.*

### 4.2. Deriving the parameter kappa from magnetic field increments

After obtaining the magnetic field increments for a given period of interest, we construct their histogram. Figure 2 presents an example of the histogram of hourly-averaged/hourly increment scale $\delta B_N$ for the first 271 days of 2023. In this study, we employ two independent techniques to determine the kappa parameters that are characteristic of the distribution of magnetic field increments: (1) Kappa fitting technique, (2) Kappa moments technique.

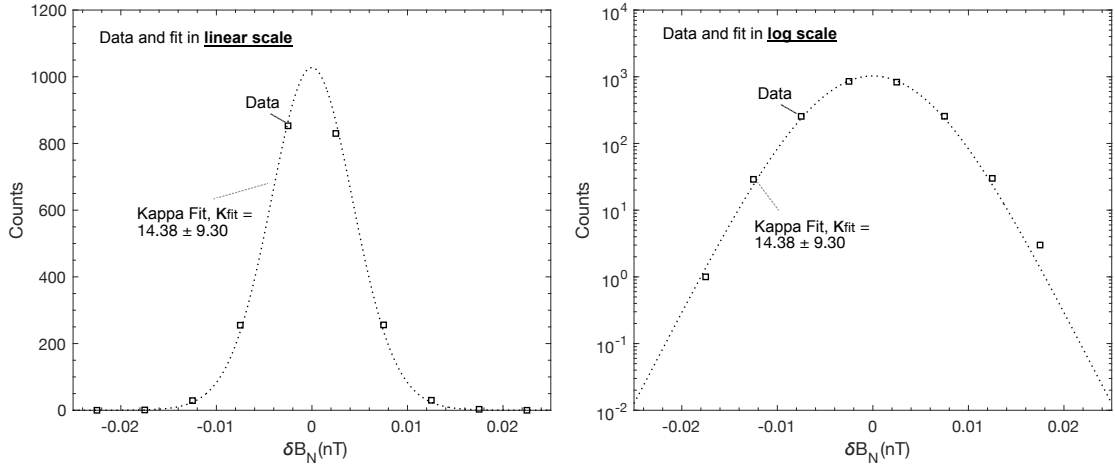

*Figure 2. Comparison of the observations and its fit presented in linear (left) and logarithmic (right) scales. The discrepancy between the observation and the fit (from the kappa fitting technique) outside of the peak region is more apparent when presenting the result in logarithmic scale. There are no counts in bins beyond -0.02 and 0.02 nT, and thus the data points are not shown when plotted in the logarithmic scale.*

### 4.2.1. Kappa Fitting technique

The kappa distribution is used for describing the statistics of a d-dimensional vector, *x*, and can be generalized as (Livadiotis & McComas 2013):

$$P(x) = (2\pi\kappa_0\sigma_x^2)^{-d/2} \cdot \frac{\Gamma(\kappa_0 + 1 + \frac{d}{2})}{\Gamma(\kappa_0 + 1)} \cdot \left[1 + \frac{1}{\kappa_0} \cdot \frac{(x-\mu_x)^2}{2\sigma_x^2}\right]^{-\kappa_0 - 1 - d/2} \tag{3}$$

where $\kappa_0$ is the invariant kappa value (Livadiotis & McComas 2011b), *d* is the dimensionality of the examine magnetic field component, (i.e., d = 1 for a single component of the total vector, d = 2 for a planar vector component, d = 3 for the total vector magnitude), $\mu_x$ is the mean value of *x*, $\sigma_x$ is the standard deviation of the magnitude of each component of the vector $(x - \mu_x)^2$, and $\Gamma$ is the gamma function. This study analyzes increments of the magnetic field components ($\delta B$), where we use Eq. 3 for the case of d = 1 and $\mu$=0. Since $\kappa$ is related to $\kappa_0$ by $\kappa = \kappa_0 + 3/2$, we can also rewrite $-\kappa_0 - \frac{3}{2}$ in Equation 3 as just -$\kappa$, as shown in Eq. 5.

The kappa fitting technique is performed by fitting the model counts, $P_{fit}(\delta B, \kappa)$, to the observed counts of the histogram of $\delta B$ (in logarithmic space) and minimizing the reduced chi-square equation:

$$\chi^2_{reduced}(\kappa) = \frac{1}{M-1} \cdot \sum_{l=1}^{M} C_l \cdot \ln^2 \frac{C_l}{P_{fit}(\delta B_{i,bin};\kappa)} \tag{4}$$

Such that:

$$P_{fit}(\delta B_{i,bin}, \kappa) = P_0(\kappa) \cdot \left[1 + \frac{(\delta B_{i,bin})^2}{(2\kappa-3)\cdot M_1}\right]^{-\kappa} \tag{5}$$

and

$$P_0(\kappa) = \Delta(\delta B) \cdot \sum_{l=1}^{M} C_l(\delta B_{i,bin}) \cdot \frac{1}{\sqrt{\pi M_1} \cdot \sqrt{2\kappa-3}} \cdot \frac{\Gamma(\kappa)}{\Gamma(\kappa-0.5)} \tag{6}$$

where

$$M_1 = \frac{1}{N} \sum_{j=1}^{N} (\delta B_i(t, \bar{\tau}_j)^2)$$

where $C_l(\delta B_{i,bin})$ denotes the histogram of $\delta B_i$ with *l* referring to each bin ($\delta B_{i,bin}$) in the histogram and i referring to the examined component of the magnetic field vector, $\Delta(\delta B)$ is the width of each bin, *N* is the total number of non-zero points ($\delta B_i(t, \bar{\tau}_j)$) and $M$ is the total number of histogram bins where counts are > 1. $M_1$ is the first moment of the increment distribution or the variance.

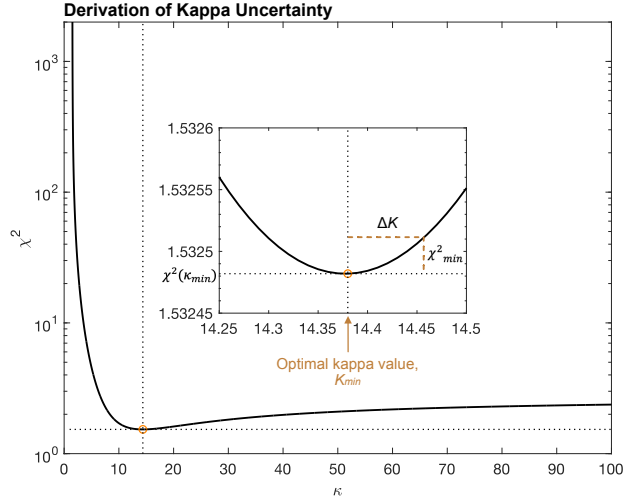

*Figure 3. Example of how the optimal kappa value and the corresponding uncertainty are determined using the kappa fitting technique.*

We compute the optimal kappa by binning the values of kappa (with a width of $\kappa$-bin equal to $\Delta\kappa = 0.1$) and finding the bin for which the reduced $\chi^2$ is minimized. In addition, we compute the uncertainty of the kappa estimate using the following equations that uses the curvature near the minimum of chi-square (Livadiotis 2007):

$$\delta\kappa = \Delta\kappa \cdot \sqrt{\frac{\chi^2_{min}}{\chi^2(\kappa_{min+\Delta\kappa}) - \chi^2(\kappa_{min})}}$$

(6)

$$\chi^2_{min} \equiv \frac{\chi^2(\kappa_{min})}{M}$$

(7)

Where $\chi^2(\kappa_{min})$ is the smallest reduced $\chi^2$, and $\chi^2(\kappa_{min+\Delta\kappa})$ is the reduced $\chi^2$ next to the smallest reduced $\chi^2$.

In this study, we perform the fittings and compute the reduced $\chi^2$ over a wide range of $\kappa$ values (from 1.5 to 100) using the equations above. An example of how the reduced $\chi^2$ varies as a function of kappa values is presented in Figure 3, along with how the uncertainty of the best kappa estimate was determined. The best $\kappa$ value is the one that minimizes the reduced $\chi^2$ value, as shown in Figure 3. In the case of no convergence, the 'optimal' kappa value is assigned to the highest kappa value (100, in this case), and its uncertainty is set to an arbitrarily large value (10000).

The fitting result for the hourly-averaged $\delta$BN from the first 271 days of 2023 is shown in Figure 2. Notably, for a given distribution that has sufficient statistics and is well-defined, the discrepancy between the data and the fit beyond the core population (near the peak) only becomes more apparent when presented in log scale. In other words, when the fit was done in linear space, the results can be biased toward a more gaussian-like fitting where the core population (which is the more gaussian-like distribution) is dominant (Nicolaou & Livadiotis 2016). This illustrates one of the motivations behind this study: to examine whether the fitting result changes when the tail of distribution is also taken into consideration by performing the fit in logarithmic space. For the rest of the study, all histogram results are shown in logarithmic scale.

### 4.2.2. *Kappa Moment Technique*

The kappa moment technique is a self-consistent way of estimating the parameter kappa empirically from the data. The technique involves the ratio of two different moments of the sample yielding a dimensionless quantity that is expressed with $\kappa_{\Delta B}$ as the only unknown variable.

The moment, $M_a$ of the increment's square is given by:

$$M_a = \left( \frac{1}{N} \sum_{j=1}^{N} |\delta B_i(t, \bar{\tau}_j)^{2a}| \right)^{1/a} \quad (8)$$

All moments are generally well defined in $a \in [0,1]$ and connected with the value of kappa through some functional (Livadiotis 2017, 2019, Ch. 5), except for a = 1 where the moment depends on the standard deviation. We choose to compute the ratio of $M_{1/2}$ to $M_1$, $g(\kappa_0)$, which gives:

$$g(\kappa_0) = \frac{M_{\frac{1}{2}}}{M_1} = \frac{2}{\pi} \cdot \kappa \cdot \left[ \frac{\Gamma(\kappa_0 + 0.5)}{\Gamma(\kappa_0 + 1)} \right]^2 \quad (9)$$

where $\kappa_0 = \kappa - 3/2$ and N is the total number of non-zero points ($\delta B_i(t, \bar{\tau}_j)$). An example of how the best kappa estimate is determined through the kappa moment technique is shown in Figure 4.

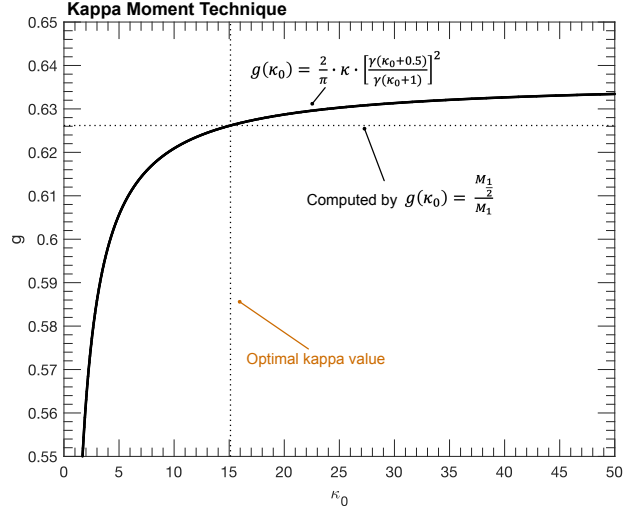

*Figure 4. An example of how the kappa moment technique is used to estimate the parameter kappa using the hourly-averaged/hourly increment scale data for $B_N$ during the first 271 days of 2023 as shown in Figure 2.*

## 5. RESULTS

This section investigates the kappa values characteristic of magnetic field increments during the first 271 days of 2023 using the approaches described in Section 4. Importantly, it explores how the kappa estimates vary for different increment windows and statistical periods.

### 5.1. *Averaged kappa values for the first 271 days of 2023*

Figure 5 shows the 1-hour magnetic field increments during the first 271 days of 2023 (consistent with Burlaga et al. 2024b). The 1-hour magnetic field increments were computed by first obtaining the hourly averaged magnetic field from the 48-second magnetic field measurements, then determining the increments using Equation 1. The magnetic field increments are binned from -0.03 to 0.03 with a bin size of 0.005 (see Appendix A for discussion of the effect of this bin width choice on the kappa results). Since the R-component has a smaller increment range, we bin $\delta B_R$ from -0.02 to 0.02 with a bin size of 0.002.

The right panels of Figure 5 present the histogram of the magnetic field increments in logarithmic scale, and the gray dotted lines show the fitted distribution from the kappa fitting technique as described in Section 4. The kappa value listed on Figure 5 is the average of kappa estimates derived from the two techniques discussed in Section 4, and the uncertainty is the difference between the two kappa estimates.

We determine that the average kappa values for all three field components are >10, which is considerably higher than what is typically observed values of thermodynamic kappa in the space plasmas in the near-sun environment (~1.5-6.6; see Livadiotis 2017, Table 1.1., and references therein). This indicates a weakening of correlations among the magnetic field increments over a long period of time such as the examined period of 271 days. The results in Figure 5 are also consistent with the results obtained by Burlaga et al. (2024b). Similar results were found for the increments of the magnetic field magnitudes for the T and N components only ($\delta B_{TN}$) and for all three components ($\delta B$). However, the choice of the increment windows and statistical periods is rather arbitrary. It remains unclear how the kappa estimates will vary if different increment windows ($\tau$ in Eq. 1) or different statistical periods are selected.

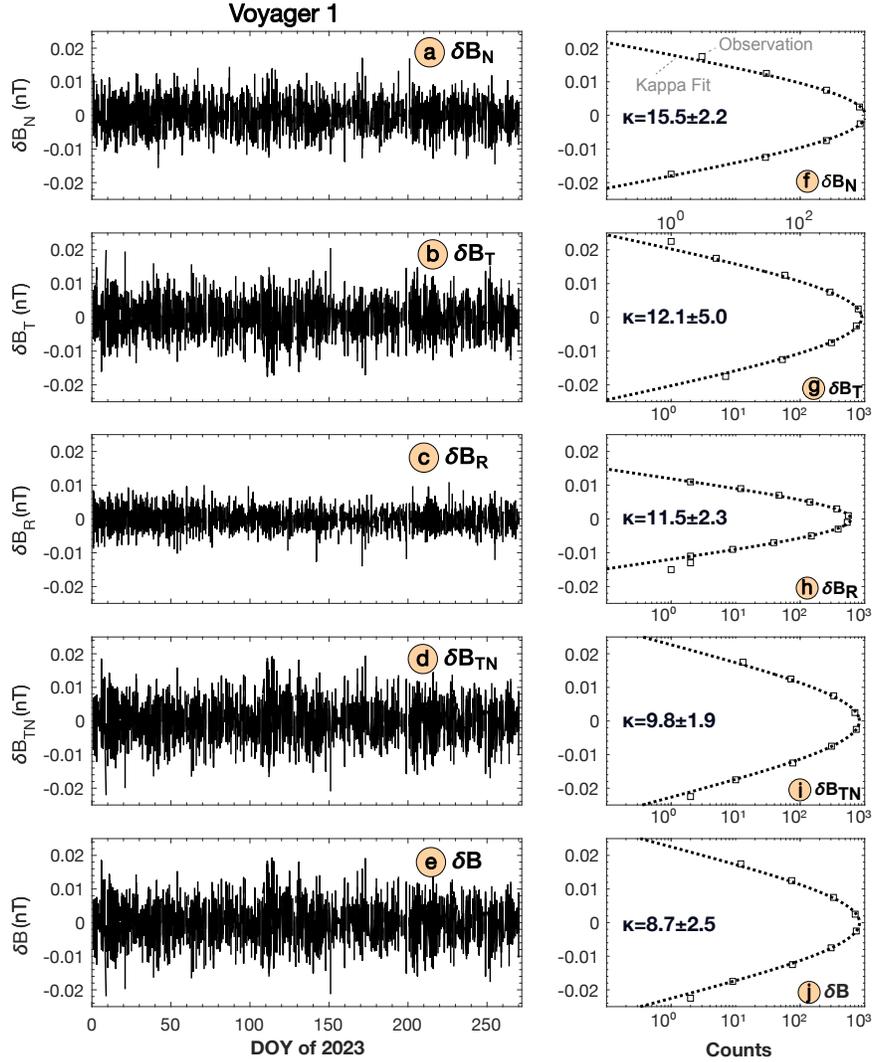

*Figure 5. (Left) Increments of magnetic field in units of nT for (a) the normal component, $\delta B_N$, (b) the tangential component, $\delta B_T$, (c) the radial component, $\delta B_R$, (d) the magnitude from the tangential and normal components, $\delta B_{TN}$ (e) the magnitude of all three field components, $\delta B$. (Right) Histogram of the magnetic field increments from the left panels for (f-j) $\delta B_N$, $\delta B_T$, $\delta B_R$, $\delta B_{TN}$, and $\delta B$, respectively.*

### 5.2. Kappa estimates with varying increment windows

Here we explore the effect of increment windows ($\tau_B$) on the kappa estimates using the kappa fitting technique (see Section 4.2.1). Figure 6 illustrates an example of how the kappa estimate, the variance of the distribution ($M_1$), and reduced $\chi^2$ values, derived from kappa fitting, change as a function of increment windows, ranging from 1 minute to 200 minutes.

All three panels show a clear trend of large fluctuations at small increment windows before approaching a plateau region beyond an increment window of 30 minutes. At shorter increment scales (< 20 minutes), the estimated kappa value appears to be close to infinity, with significant uncertainty and approaches the plateau when it reaches 30 minutes and beyond, suggesting that the kappa estimate is stable. Notably, at larger increment windows, the $M_1$ value shows a gradually increasing trend, which is expected. When we average over a wider $\Delta T$ window (see Figure 1), the variance of the distribution, as indicated by $M_1$, increases as the likelihood of including different systems into the same distribution increases.

When examining the changes in the kappa values on a smaller scale (from one increment window to the next), as shown in the zoomed-in plots in Figure 6, we observe a large variation in the kappa estimate from one window to the next. For instance, even though the kappa value is ~15 when using an increment duration of 60 minutes (as illustrated in Figure 6), the estimate changes to ~ 6 if a duration of 62 minutes is used instead. The high sensitivity of the kappa estimates on the increment window choice suggests that a more statistical approach is needed for an accurate determination of kappa.

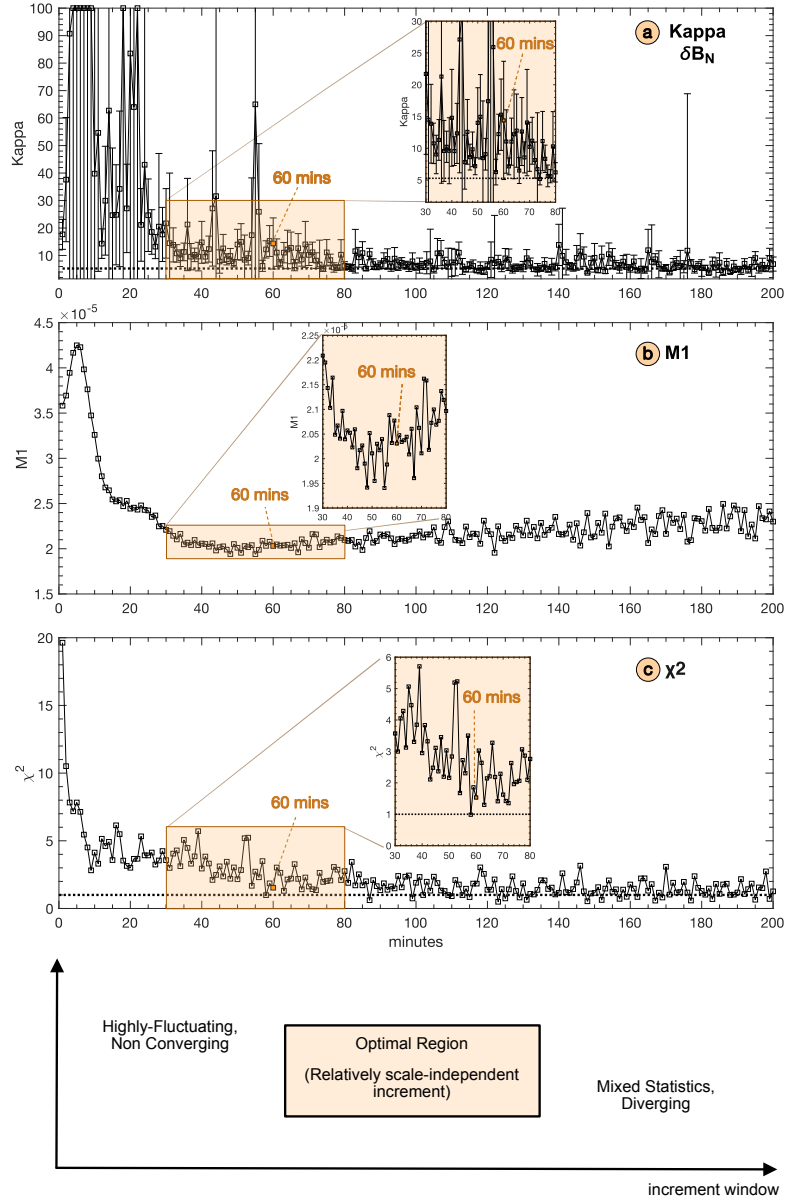

*Figure 6. Kappa values, M1, and the reduced $\chi^2$ values for the histogram of the magnetic field increments, $\delta B_N$, as a function of increment windows. The statistical period is still the first 271 days of 2023 for consistency with Section 5.1. The bottom panel provides an illustration of the kappa variations as a function of increment scale.*

Instead of relying on results derived from a single increment window, we propose the use of a weighted average of kappa values over a range of increment windows. As shown in Figure 6, an increment window anywhere between 30 and 100 minutes is a reasonable choice because both the kappa and $M_1$ values stabilize to a plateau. Therefore, the weighted average of kappa values corresponding to the windows from 30 to 100 minutes is used for the rest of this study; in several cases (e.g., Figure 7), we use even larger resampling intervals, because individual values of kappa and $M_1$ appear to be stabilized beyond 100 minutes, even up to 200 minutes. The weighted-average kappa is computed with the variance of the kappa values to be the weighting factor; in other words, values of kappa with larger uncertainty will contribute less to the average kappa than values with smaller uncertainty. We note that although this approach chooses to average over a period where kappa values are relatively stable (or largely independent of the increment window/scale, as it approaches the plateau) and neglects the highly fluctuating kappa values from the finer increment scale, it is still possible that there is smaller-scale intermittency within finer increment scales (e.g., Fraternale et al. 2019, 2021).

Figure 7 shows the kappa values for $\delta B_N$, $\delta B_T$, $\delta B_R$, $\delta B_{TN}$, and $\delta B$ as a function of increment windows and highlights the weighted-average kappa value (derived from 1-200 minutes increment windows) with purple lines. The right panels of Figure 7 present the histogram and the enriched histogram (black and blue lines) of the kappa values from the left panels. The enriched histogram is computed by randomly generating kappa values (in this case, 10 points for each increment window) from normal distributions. The mean of each distribution is the estimated kappa value of that window, and the standard deviation is the corresponding kappa uncertainty. Notably, the average kappa values (purple lines) for $\delta B_N$, $\delta B_T$, $\delta B_R$, $\delta B_{TN}$, and $\delta B$ generally coincide with the peak of the two histograms, suggesting that the weighted-average kappa values are representative of the period we intended to characterize without biasing to a specific choice of increment window.

In addition, the weighted-average kappa values for all magnetic field increments are lower (varying between $\sim$ 4–6) than those found in Figure 7 ($\sim$9 – 16). This implies that kappa falls within a range that is not approximated by a Gaussian when we consider the variation of the estimates as a function of increment windows in the analysis. This also leads to our next question: how does the statistical period affect the kappa?

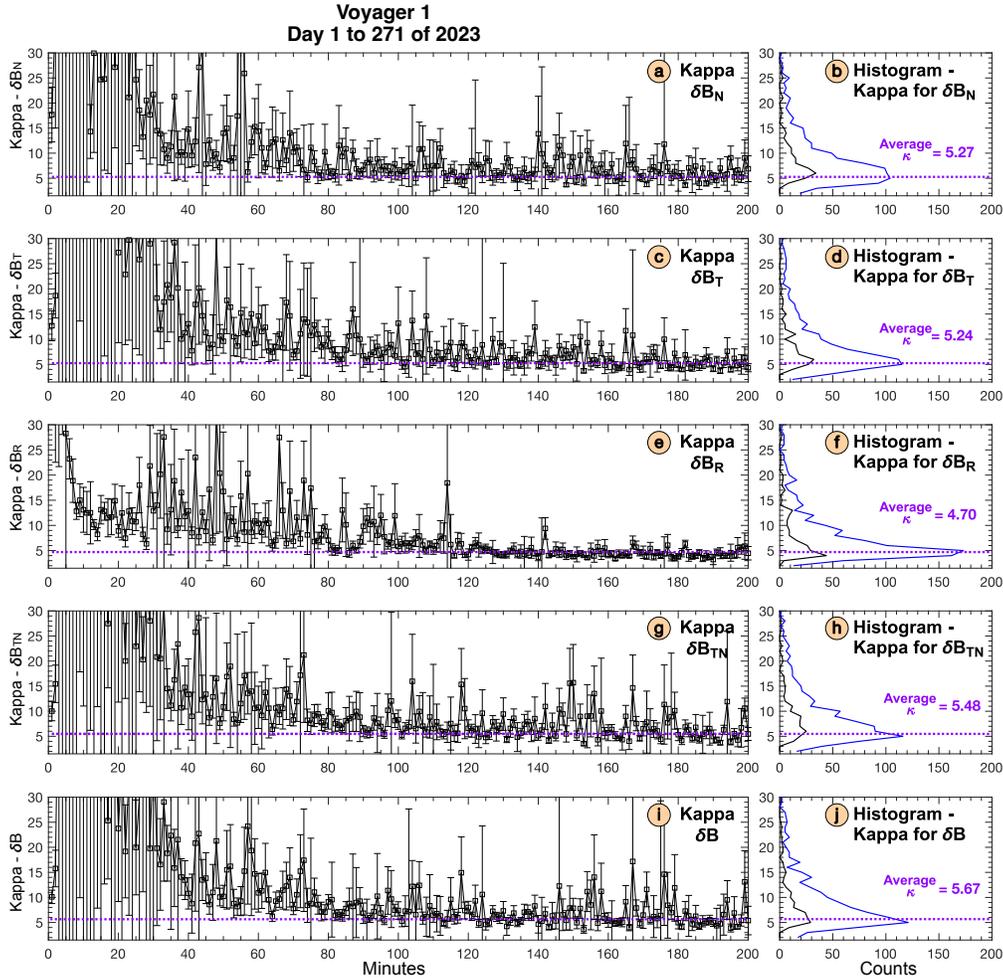

*Figure 7. (Left) Similar to Figure 5, but for the kappa values and its corresponding uncertainty at each increment window, ranging from 1 minutes to 200 minutes. (Right) Histogram and enriched histogram of the kappa values from the left panels for $\delta B_N$, $\delta B_T$, $\delta B_R$, $\delta B_{TN}$, and $\delta B$.*

### 5.3. Average kappa estimates with varying statistical windows

We further investigate how the choice of statistical period, which has been the same in the previous sections (the first 271 days of 2023), affects the kappa estimates. To study this, we examine the variation of the weighted-average kappa for two cases: (a) centered at Day 136 and, with an increment of 6 days for each step (3 days on each side), gradually increasing the length of the statistical periods until it covers the entire 271 days; (b) non-overlapping 30-day statistical periods in 2023 such that it shows the average kappa estimate for the first 30 days of 2023, then the subsequent 30-day periods until it covers the entire interval. We perform the same analysis, as described in Section 5.2 (see also Figures 6 and 7), and average the kappa values over the plateau region of the increment windows (30-100 minutes).

Figure 8 presents the weighted average kappa values for different magnetic field components and magnitudes. We find that the average kappa values are low for <20 days. The kappa estimates increase after day ∼20 and remain relatively stable between ∼30-110 days with a gradual upward trend thereafter (see Figure 8a). The low kappa estimates for shorter statistical periods like <20 days are likely due to poor statistics, as discussed in Appendix B. We further examine how kappa estimates vary over a 30-day non-overlapping interval in 2023 (Figure 8b). In general, the average kappa values are quite stable and are consistently within 3-7 for all magnetic field components. The increase in kappa values for longer statistical periods—such as those beyond 110 days—suggests that different regions or structures may be combined into a single distribution when the statistical window is too large, leading to an overestimation of the kappa value.

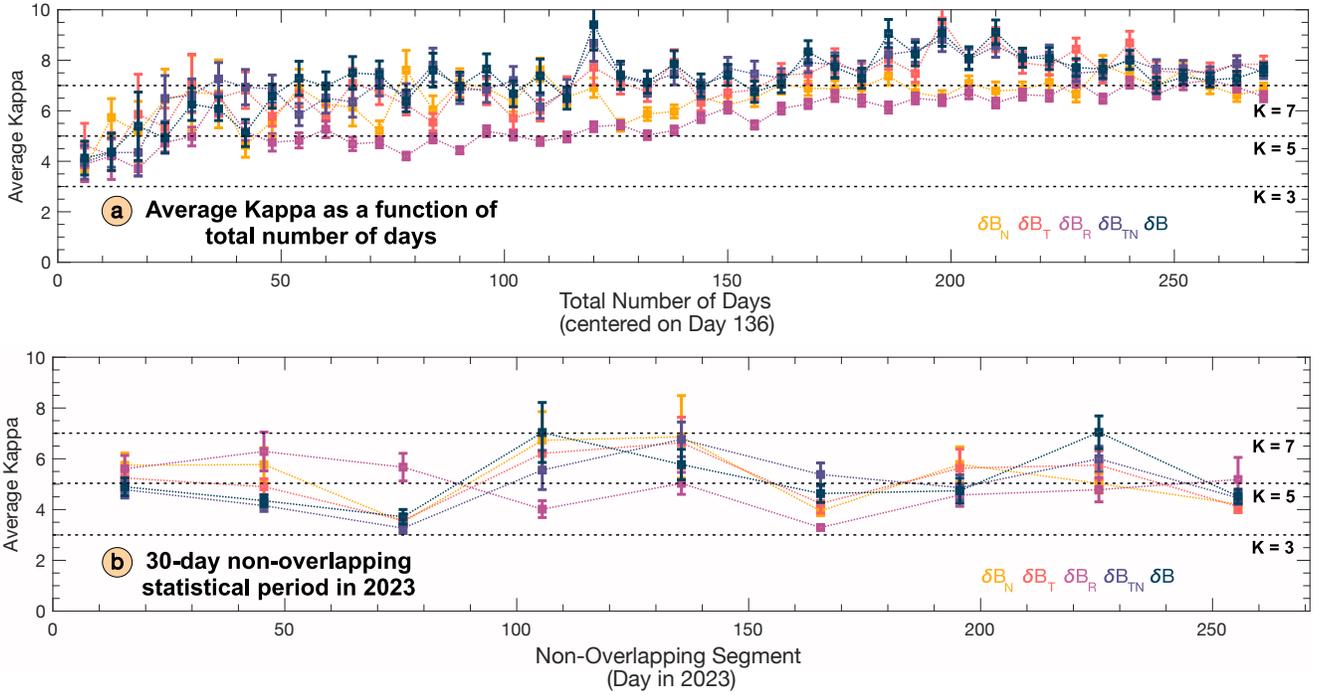

*Figure 8. Variations of weighted-average kappa values (a) as a function of total statistical periods (total number of days included for the analysis) and for non-overlapping 30-day statistical periods in 2023. Both panels eventually cover the same period of days (Days 1 to 271) in 2023 for consistency purposes.*

## 6. DISCUSSION AND SUMMARY

This study investigates how the choice of increment windows and the statistical period affect the fitted distribution of magnetic field increments. The fitting was performed in logarithmic space, which improves the inclusion of the distributions' tails. We note that for 1-hour magnetic field increments over the first 271 days of 2023, the kappa values that were derived from two different kappa techniques are high (~11 with good agreements between the two kappa methods) across all magnetic field components. This is generally consistent with the findings from Burlaga et al. (2024b) where the magnetic field increments during this period are close to Gaussian.

Our further analysis, however, indicates that the kappa values estimated by the fitted distribution of the magnetic field increments is highly sensitive to the choice of the increment window - the value can vary significantly from one increment window to another. This is likely a result of the high fluctuation in the average data due to data gaps in Voyager-1 measurements. Therefore, we use average kappa values over a broad increment window range, as the average kappa value is found to coincide with (or near) the peak of the kappa histogram as illustrated in Figure 7.

Overall, we determine that the average kappa values for $\delta B_N$, $\delta B_T$, $\delta B_R$, $\delta B_{TN}$, and $\delta B$ are between 4.70 and 5.67 for the same statistical period (the first 271 days of 2023). Particularly, $\delta B_N$ and $\delta B_T$ have very similar average kappa values, ~5.25. Similarly, $\delta B_{TN}$ and $\delta B$ have their average kappa values, ~5.56. Meanwhile, $\delta B_R$ seems to have the smallest average kappa value, ~4.7. These values are higher compared to other regions of the heliosphere, such as the heliosheath (between 1.5 and 2.5; Livadiotis & McComas 2013) but are not close to being a Gaussian distribution as suggested by Burlaga et al. (2024b).

The kappa value also varies significantly as a function of increment windows. Smaller increment windows correspond to larger estimation of kappa, such that the distributions appear nearly Gaussian. The kappa value plateaus for an increment window of 30-minute or larger. This implies that an increment window of <30 minutes might be too small to detect coherent structures in the VLISM or the presence of coherent structures are limited to scales larger than 30 minutes. It is also important to note that the common practice in intermittency studies is to use an increment scale that is larger than the resampling scale [e.g., Fraternale et al. 2022]. Ideally, one should use the highest resolution data for such a study. In this study, we used the sequential magnetic field increment where the increment window is the same as resampling window to remain consistent with the method used by Burlaga et al. (2024b); this allows us to focus on understanding the effect of varying resampling window and statistical window on the result. We note that resampling could reduce the effect of data gap and suppress the data noise (1 sigma uncertainty for $B_N$, $B_T$, and $B_R$ are 0.02, 0.02, and 0.06 nT, respectively), but it could also alter/suppress intermittency. Further analysis could explore the effect of increment scale on the intermittency.

When we examine the changes in kappa estimates under varying statistical analysis periods, we find a rather constant average kappa estimate (within 3-7) for all magnetic field components when using a 30-day statistical period. The kappa value increases as we increase the statistical analysis period. We find that the small kappa values associated with short statistical periods like *<20* days are likely due to poor statistics. In contrast, a 30-day period appears to be a reasonable choice for investigating magnetic field increments via this methodology where the kappa values derived from statistical periods ranging from 30 to ~110 days remain relatively stable. For periods longer than 110 days, the kappa values gradually increase, which may result from the inclusion of statistics from other systems, thereby yielding a statistically induced, Gaussian-like result.

Importantly, the low average kappa estimates indicate that Voyager 1 is still in an intermittent magnetic field environment almost 11 years after the generally accepted heliopause location (Gurnett et al. 2013). This result contradicts the finding of Burlaga et al. (2024b) and indicates that Voyager has not entered a new region nor just crossed the heliopause as predicted by Fisk & Gloeckler (2022). As noted by Fraternale et al. 2022, "the presence of intermittency and scaling laws are key ingredients of turbulence". The low kappa estimates also suggest that there exist some coherent structures within the VLISM where Voyager 1 was with the indication of shocks/waves and turbulence within the VLISM, possibly emanating from the heliosphere (e.g., Fraternale et al. 2020; Zank et al. 2019).

In summary, this work probes at the question of whether and how the magnetic field increments in the VLSIM follows the gaussian distribution by introducing a new, statistically based framework to analyze magnetic field increments. Our analysis leads to a different conclusion than previous findings and indicates that the magnetic field increments in the VLISM is not Gaussian and exhibits intermittency (low kappa values, ~3-7). This work also calls for a more careful approach when conducting such an analysis where an arbitrary choice of increment windows or statistical periods could give biased results. Although this work only uses Voyager 1 data from the VLISM, this approach is applicable for any other measurements of the magnetic field in any astrophysical environment. The framework that we present here will serve as the cornerstone for future studies to address important questions such as how intermittency evolves from the outer heliosphere into the interstellar medium.

We are grateful to everyone who helped make the Voyager mission and their data possible. This work is funded in part by the IMAP mission as a part of NASA's Solar Terrestrial Probes (STP) Program (80GSFC19C0027). The authors also would like to thank Eric Zirnstein for helpful discussions on this work.

## Appendix A: Effect of the Bin Width on Weighted-Averaged Kappa

This appendix explores the effect of bin width—or histogram resolution—on the weighted-average kappa. Figure A1 shows the histogram for a 6-day, 18-day, 60-day, and 180-day statistical period (centered on Day 137 of 2023) using a $\Delta dB$ bin size from 0.005 nT (which is nominally used for $dB_N$, $dB_T$, $dB$, and $dB_{TN}$; top panels) and 0.002 nT (bottom panels). We find that the histograms retain the overall shape of the distribution for both bin width choices.

To further evaluate the effect of histogram resolution on the kappa fitting result, we repeated the analysis in Section 5.3 using 0.002 nT. These results are shown as red points in Figure A2, which displays the weighted average of kappa values over 30-100 minutes for the selected statistical periods. For varying statistical periods, the kappa values derived from the finer $\Delta dB$ bin sizes are generally consistent, though slightly lower than those from the larger $\Delta dB$ bin sizes. This indicates that a 0.005 nT bin size is adequate for the analysis and does not introduce significant bias in the weighted-average kappa, although it may slightly overestimate the values.

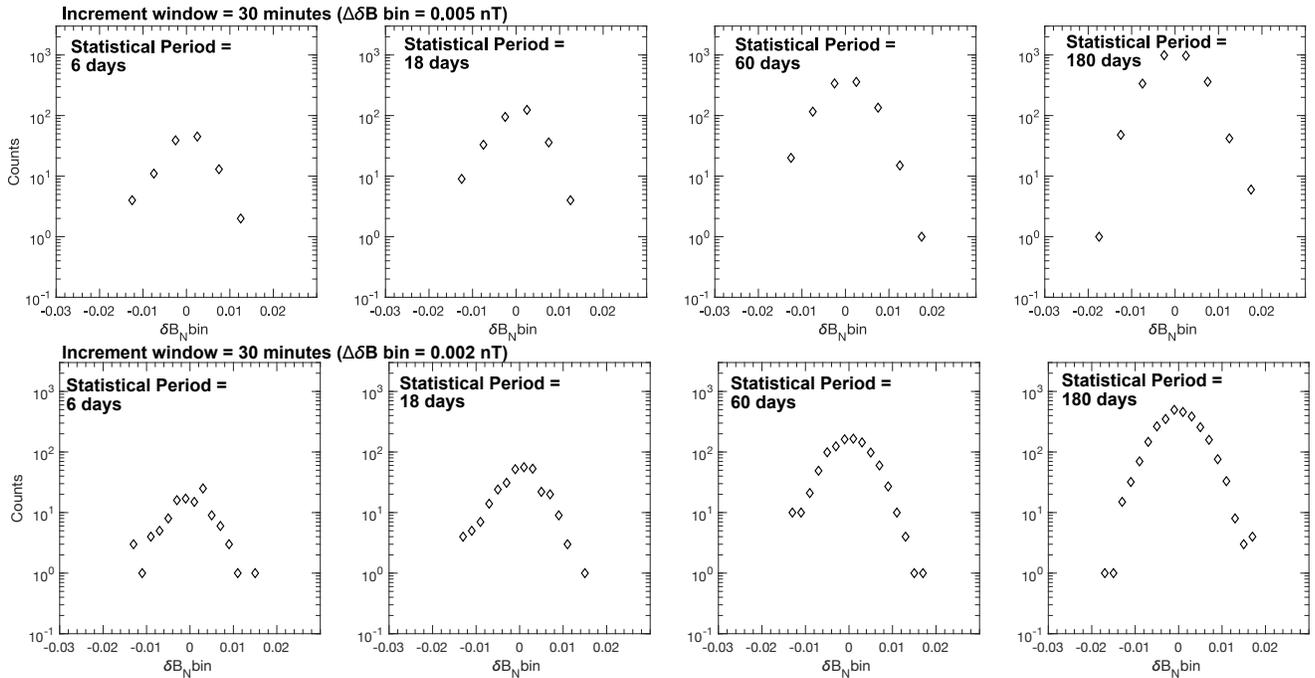

*Figure A1. Examples of histogram for $\delta B_N$ with an increment window of 30 minutes for a statistical period of 6-day, 18-day, 60-day, and 180-day, centered on Day 136 of 2023. Top panels use a $\Delta dB$ bin of 0.005 nT (similar to what was used in the manuscript for $\delta B_N$), while bottom panels use a $\Delta dB$ bin of 0.002 nT.*

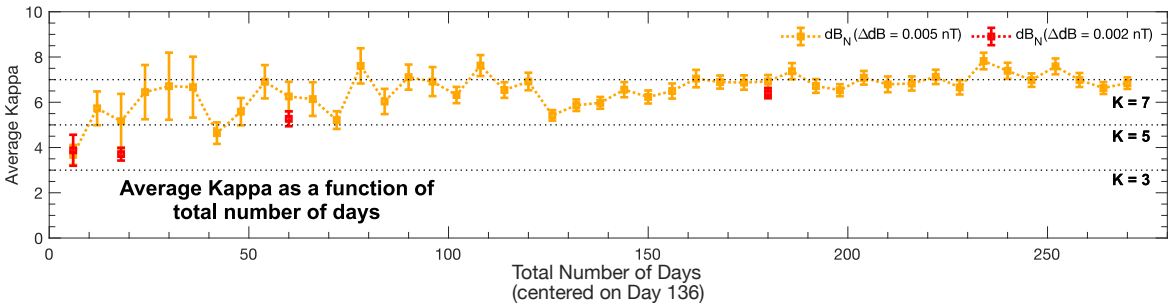

*Figure A2. Similar to Figure 8a in the manuscript but with only $\delta B_N$ data. Red points represent results derived with a finer $\Delta dB$ bin (0.002 nT).*

## Appendix B: Effect of statistics on kappa estimates

This appendix investigates the effect of statistics on kappa estimates. Figure B1 shows the number of events that have >5 sampling sizes (blue) and 2-5 sampling sizes (red) in each corresponding $\delta B_N$ bin for an increment scale of 60-minute and varying statistical periods and $\Delta \delta B$ bins. For shorter statistical period like 7 days, only two $\delta B_N$ bins consistently have >5 counts in each bin, suggesting that the statistic in the distribution is likely insufficient to properly characterize the increments. The statistics improve with an inclusion of longer statistical periods such as 20 days and 30 days. Figure B2 further demonstrates the influence of poor statistics on the kappa estimates. As expected, the shorter statistical period that has poor statistics generally has lower kappa estimates. For those with longer statistical periods (such as those between 20-45 days), the kappa estimates are consistent with each other, between 3 to 7.

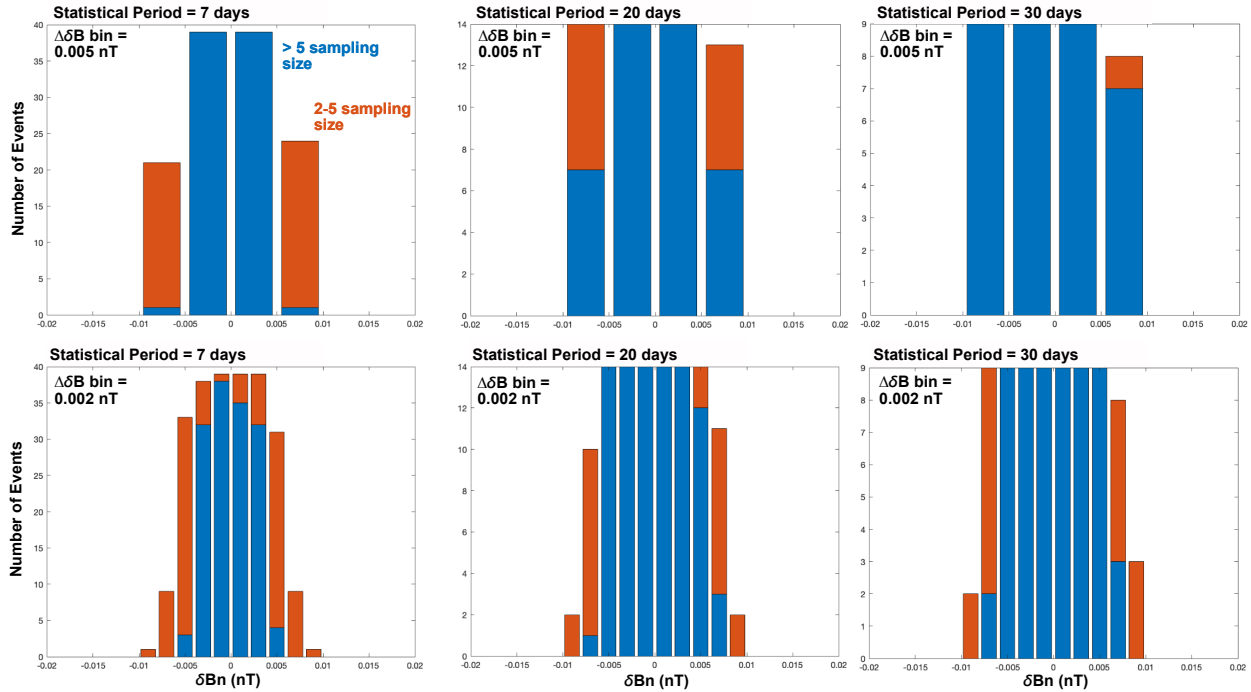

*Figure B1. Histogram of $\delta B_N$ at an increment scale of 60-minute for three statistical periods: 7 days, 20 days, and 30 days with bin widths of 0.005 nT (top panels) and 0.002 nT (bottom panels). It indicates the number of events with a sampling size greater than 5 (blue) and a sampling size of 2-5 (red).*

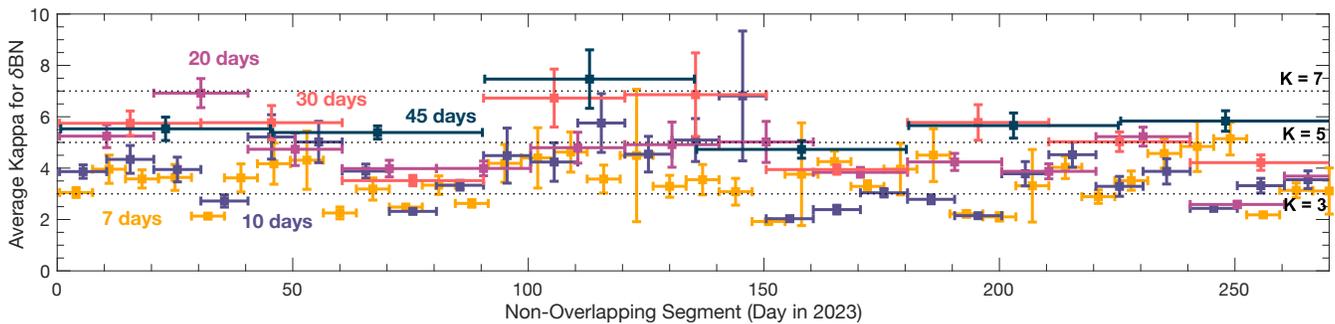

*Figure B2. Like Figure 8a but for varying statistical periods (7, 10, 20, 30, and 45 days) and only for $\delta B_N$.*